\documentclass[preprint,showpacs,preprintnumbers,amsmath,amssymb,graphicx]{revtex4}
\usepackage{amsmath,amsfonts,amssymb}
\usepackage{epsfig}
\usepackage{color}
\def\bit{\begin{itemize}}
\def\eit{\end{itemize}}
\def\ben{\begin{enumerate}}
\def\een{\end{enumerate}}
\def\bed{\begin{description}}
\def\eed{\end{description}}

\def\k{\kappa}
\def\l{\lambda}

\def\lsim{\raise0.3ex\hbox{$<$\kern-0.75em\raise-1.1ex\hbox{$\sim$}}}
\def\gsim{\raise0.3ex\hbox{$>$\kern-0.75em\raise-1.1ex\hbox{$\sim$}}}

\let\jnfont=\rm
\def\NPB#1,{{\jnfont Nucl.\ Phys.\ B}{\bf #1},}
\def\PLB#1,{{\jnfont Phys.\ Lett.\ B}{\bf #1},}
\def\EPJC#1,{{\jnfont Eur.\ Phys.\ Jour.\ C}{\bf #1},}
\def\PRD#1,{{\jnfont Phys.\ Rev.\ D}{\bf #1},}
\def\PRL#1,{{\jnfont Phys.\ Rev.\ Lett.\ }{\bf #1},}
\def\MPLA#1,{{\jnfont Mod.\ Phys.\ Lett.\ A}{\bf #1},}
\def\JPG#1,{{\jnfont J.\ Phys.\ G}{\bf #1},}
\def\CTP#1,{{\jnfont Commun.\ Theor.\ Phys.\ }{\bf #1},}
\def\JHEP#1,{{\jnfont JHEP }{\bf #1},}
\def\NPPS#1,{{\jnfont Nucl.\ Phys.\ Proc.\ Suppl.\ }{\bf #1},}

\newcommand{\beq}{\begin{eqnarray}}
\newcommand{\eeq}{\end{eqnarray}}
\newcommand{\bpmatrix}{\begin{pmatrix}}
\newcommand{\epmatrix}{\end{pmatrix}}
\newcommand{\fr}{\frac}
\newcommand{\la}{\lambda}
\newcommand{\crn}{\nonumber \\}
\newcommand{\ep}{\epsilon}
\newcommand{\hc}{\text{ h.c}}

\newcommand{\diag}{\text{diag}}

\newcommand{\ba}{\begin{array}}
\newcommand{\ea}{\end{array}}

\newcommand{\be}{\begin{equation}}
\newcommand{\ee}{\end{equation}}

\begin{document}
\title{Probing degenerate heavy Higgs bosons in NMSSM with vector-like particles}

%\author{Ning Liu $^1$}
\author{Fei Wang $^2$}
\author{Wenyu Wang$^3$}
\author{Lei Wu$^1$}\email{leiwu@itp.ac.cn}
\author{Jin Min Yang$^4$}
\author{Mengchao Zhang$^4$}

\affiliation{
$^1$ Department of Physics and Institute of Theoretical Physics, Nanjing Normal University, Nanjing, Jiangsu 210023, China\\
$^2$ School of Physics, Zhengzhou University, Zhengzhou 453000, China\\
$^3$ Institute of Theoretical Physics, College of Applied Science,
     Beijing University of Technology, Beijing 100124, China\\
$^4$ Key Laboratory of Theoretical Physics, Institute of Theoretical Physics, Chinese Academy of Sciences,
              Beijing 100190, China}

\begin{abstract}
In this work, we investigate the degenerate heavy Higgs bosons in the Next-to-Minimal Supersymmetric Standard Model (NMSSM)
by introducing vector-like particles. Such an extension is well motivated
from the top-down view since some grand unified theories usually predict the existence of singlet
scalars and vector-like particles at weak scale. Under the constraints from the LHC and dark matter experiments,
we find that (1) the null results of searching for high mass resonances have tightly constrained the parameter space; (2) two degenerate heavy singlet Higgs bosons $h_2$ and $a_1$ can sizably decay to $\chi^0_1\chi^0_1$ invisibly. Therefore, search for the monojet events through the process $gg \to h_2 / a_1 (\to \chi^0_1 \chi^0_1) j$ may further test our scenario at the future LHC.
\end{abstract}
\maketitle
\section{Introduction}
Weak-scale supersymmetry (SUSY) resolves several problems in the Standard Model (SM), most notable of which
is the gauge hierarchy problem \cite{susyhierarchy}, i.e. the difference between the electroweak scale and
the Planck mass. A consequence of this theory is that the supersymmetric particles and the Higgs bosons
should exist with masses typically less than 1 TeV. In fact, the discovery of the 125 GeV Higgs
boson \cite{higgs-atlas,higgs-cms} may be the first evidence of supersymmetry since it lies miraculously in
the Higgs mass window $115-135$ GeV predicted by the Minimal Supersymmetric Standard Model (MSSM) \cite{mssm}.
So, if SUSY is indeed the new physics beyond the SM, more new particles could be found at the running or future LHC.

In fact, vector-like extensions are common in model buildings from the top-down approach. For example, vector-like
particles can appear in certain GUT models, in extra-dimensional models or from the dimension deconstruction.
The vector-like particles can also play an important role in SUSY breaking. For example, in some popular SUSY breaking
models, such as SUGRA or gauge mediation supersymmetry breaking, the tree-level relation $m_{H_u}^2=m_{H_d}^2$ will always
be predicted. The electroweak symmetry breaking has to be driven by quantum corrections with such boundary conditions.
In general, the large negative contributions to $m^2_{H_u}$ from the renormalization group equation (RGE) are important.
If the boundary scale is low, the additional contributions from the heavy quarks to $m^2_{H_u}$ is welcome to successfully
trigger the EWSB. The vector-like particles, which will not spoil the chiral structure of the MSSM, are the most natural
extensions. Similar reasons hold for the NMSSM. So, we introduce vector-like particles in the NMSSM and allow them to
couple with the singlet field.

In the NMSSM \cite{nmssm}, the singlet field $S$ is introduced to account for the notorious $\mu$-problem \cite{muew}.
After the EWSB, two singlet-like Higgs bosons (one is CP-even, the other is CP-odd) are obtained in the limit of
small $\lambda$. Since the two singlet-like Higgs bosons are usually related, they may become degeneracy in certain limit, which has not been studied in the literatures. In this work, we investigate the degenerate heavy singlet-like Higgs bosons
in the NMSSM with vector-like particles in the LHC and dark matter experiments. This paper is organized as follows. In Sec. \ref{sec2}, we extend the NMSSM by introducing vector-like particles. In Sec. \ref{sec3} we scan the parameter space and present the numerical
results. Finally, the conculsion is given in Sec. \ref{sec4}.

\section{NMSSM with vector-like particles}\label{sec2}
In the NMSSM, a singlet field $S$ with coupling $S H_u H_d$ is introduced and the $\mu$-term is dynamically generated
when $S$ develops a vacuum expectation value (vev). In addition, the little hierarchy problem of the MSSM can be relaxed
by the extra tree-level contributions to the SM-like Higgs boson mass. The Higgs superpotential of the NMSSM is given
by \cite{nmssm},
\beq\label{2.1e}
W_\mathrm{NMSSM} = \lambda \widehat{S}\,\widehat{H}_u \cdot
\widehat{H}_d + \frac{\kappa}{3} \widehat{S}^3
\eeq
where $\l$ and $\k$ are dimensionless parameters. Then, a vev $s$ of $\widehat{S}$ of the order of the weak or
SUSY breaking scale generates an effective $\mu$-term with
\beq\label{2.7e}
\mu_\mathrm{eff} = \l v_s\; ,
\eeq
which solves the $\mu$-problem of the MSSM. The full tree-level Higgs potential can be written as
\beq
V^{(0,NMSSM)} &=& (|\la S|^2 + m_{H_u}^2)H_{u}^\dagger H_{u} + (|\la S|^2 + m_{H_d}^2)H_{d}^\dagger H_{d} +m_S^2 |S|^2 \crn
&& + \fr18 (g_2^2+g_1^{2})( H_{u}^\dagger H_{u}-H_{d}^\dagger H_{d} )^2
+\fr12g_2^2|H_{u}^\dagger H_{d}|^2\crn
&&   + |\ep^{\alpha\beta} \la  H^{\alpha}_{u}  H^{\beta}_{d} + \kappa S^2 |^2+
\big[\ep^{\alpha\beta}\la A_\la H^{\alpha}_{u}  H^{\beta}_{d} S  +\fr13 \kappa
A_{\kappa} S^3+\hc \big] \,,
\label{nmssmp}
\eeq
where $A_{\lambda}$ and $A_{\kappa}$ are the corresponding trilinear soft breaking parameters. To clearly see the properties
of the Higgs sector, we expand the neutral scalar fields around the vevs as
\begin{eqnarray}
{\rm Re} \, H_d^0 &=& (v_d- H \sin\beta +h \cos\beta)/\sqrt{2}, \quad {\rm Im}\, H_d^0= (P \sin \beta + G^0 \cos \beta)/\sqrt{2}, \nonumber \\
{\rm Re} \, H_u^0 &=& (v_u+ H \cos\beta +h \sin\beta)/\sqrt{2}, \quad {\rm Im}\, H_u^0= (P \cos \beta - G^0 \sin \beta)/\sqrt{2}, \nonumber \\
{\rm Re} \, S &=& (v_s+s)/\sqrt{2}, \quad \quad \quad \quad \quad \quad \quad \quad
{\rm Im}\, S= P_S/\sqrt{2}.
\label{nmssm-higgs}
\end{eqnarray}
Substituting Eq.~(\ref{nmssm-higgs}) into Eq.~(\ref{nmssmp}), we can obtain the tree-level mass matrix squared
$M_S^2$ for the neutral Higgs bosons as
\begin{align} M^2 =
 \fr 12 \bpmatrix H,h,s \epmatrix  {M_S^2}
\bpmatrix  H\\h\\s \epmatrix +  \fr 12 \bpmatrix P,P_s \epmatrix  {M_P^2}
\bpmatrix  P\\P_s \epmatrix\,.
\label{eq:vhpot}
\end{align}
The tree-level $M_{S\,ij}^2$ and $M_{P\,ij}^2$ are given by \cite{roman}
\begin{eqnarray}\label{mass1}
M_{S\,11}^2&=& M_A^2 + (M_Z^2- \frac{1}{2} \lambda^2 v^2) \sin^2 2 \beta,\\
M_{S\,12}^2&=&- \frac{1}{2} (M_Z^2- \frac{1}{2} \lambda^2 v^2) \sin 4 \beta, \\
M_{S\,13}^2&=&-\sqrt{2}\lambda v\mu x \cot2\beta,\\
M_{S\,22}^2&=&M_Z^2\cos^2 2 \beta + \frac{1}{2} \lambda^2 v^2 \sin^2 2 \beta,\label{sm-like}\\
M_{S\,23}^2&=&\sqrt{2}\lambda v\mu (1-x),\\
M_{S\,33}^2&=&4\frac{\kappa^2}{\lambda^2}\mu^2+\frac{\kappa}{\lambda}A_{\kappa}\mu+\frac{\lambda^2v^2}{2}x-\frac{\kappa\lambda}{2}v^2\sin2\beta,\\
M_{P\,11}^2&=& M_A^2,\\
M_{P\,12}^2&=& \frac{1}{2} (M_A^2\sin2\beta-3 \lambda \k v_s^2) v/v_s, \\
M_{P\,22}^2&=&\frac{1}{4}(M^2_A\sin2\beta+3\l\k v_s^2) v^2/v^2_s\sin2\beta-3\k v_s A_\k/\sqrt{2},
\end{eqnarray}
with
\begin{eqnarray}
M_A^2 &=& \frac{\lambda v_s}{\sin 2 \beta} \left( \sqrt{2} A_{\lambda} + \kappa v_s \right), \quad x = \frac{1}{2\mu}(A_\lambda+2\frac{\kappa}{\lambda}\mu).
\end{eqnarray}
Here, it should be noted that the mass parameter $M_A$ in the NMSSM can become the mass of the pseudoscalar Higgs boson only in the
MSSM limit ($\lambda, \kappa \to 0$ with the ratio $\kappa/\lambda$ fixed). In the NMSSM, $M_A$ can be traded by the soft parameter
$A_\lambda$.

The CP-even and CP-odd Higgs mass eigenstates $h_i$ ($i=1,2,3$) and $a_i$ ($i=1,2$) can be respectively obtained by diagonalizing
${M_S^2}$ and ${M_P^2}$ with the rotation matrices ${\cal O}$ and ${\cal O}'$:
\beq
 h_i &=&  {\cal O}_{i\alpha}h_{\alpha},~~ (h_\alpha=H,h,s), \quad \diag(m_{h_1}^2,m_{h_2}^2,m_{h_3}^2)= {\cal O} {M_S^2} {\cal O}^T \; \nonumber\\
 a_i &=& {\cal O'}_{i\alpha}P_{\alpha},~~ (P_\alpha=P,P_s), \quad \diag(m_{a_1}^2,m_{a_2}^2)= {\cal O^{'}} {M_P^2} {\cal O^{'}}^T
\eeq
Here the elements of the rotation matrices satisfy the following sum rules
\begin{eqnarray}
{\cal O}^2_{1\alpha} + {\cal O}^2_{2\alpha} + {\cal O}^2_{3\alpha}&=&1,\nonumber \\
{\cal O'}^2_{1\alpha} + {\cal O'}^2_{2\alpha}&=&1 .
\end{eqnarray}
The mass eigenstates $h_{i}$ and $a_i$ are aligned by the masses $m_{h_1} \le m_{h_2} \le m_{h_3}$ and $m_{a_1} \le m_{a_2}$, respectively.
The singlet components in a physical Higgs boson $h_i$ ($a_i$) are determined by the rotation matrix
elements ${\cal O}_{is}$ (${\cal O'}_{is}$).

%%%%Table 1 %%%%%%%%%%%%%%%%%%%%%%
\begin{table}[h]
\centering
\begin{tabular}{c|ccc}
      &SU(3) &  SU(2)&  U(1) \\
\hline
 $\widehat{X_t}$ & $\bar{3}$ & $1$ & $-\frac{2}{3}$ \\
 $\widehat{Y_t}$ & $3$       & $1$ & $ \frac{2}{3}$ \\
 $\widehat{X_b}$ & $\bar{3}$ & $1$ & $ \frac{1}{3}$ \\
 $\widehat{Y_b}$ & $3$       & $1$ & $-\frac{1}{3}$ \\
 $\widehat{X_l}$ & $1$       & $1$ & $   1        $ \\
 $\widehat{Y_l}$ & $1$       & $1$ & $  -1        $
\end{tabular}
\caption{The quantum number of the vector particles under the SM gauge group. }
\label{tab2}
\end{table}

We introduce the vector-like top, bottom, and lepton multiplets $X_{ti}$/$Y_{ti}$, $X_{bi}$/$Y_{bi}$, and $X_{li}$/$Y_{li}$ with the gauge symmetry in Table \ref{tab2},
where the subscript $i$ denotes the generation of vector multiplets.
In order to forbid the mixing between vector fermion and SM fermion, we also introduce a discrete $\mathbf{Z_2}$ parity, with vector particles carry $-1$ and others carry $1$.
Then, the $\mathbf{Z_3}$ invariant superpotential of the vector
multiplets is
\begin{eqnarray}
  \label{eq:sp-xy}
W_{\rm vector}=\sum_{i=1}^n \left( \lambda_{V_{ti}}\widehat{S}\widehat{X_{ti}}\widehat{Y_{ti}}
+\lambda_{V_{bi}}\widehat{S}\widehat{X_{bi}}\widehat{Y_{bi}}
+\lambda_{V_{li}}\widehat{S}\widehat{X_{li}}\widehat{Y_{li}}   \right) .
\end{eqnarray}
All other terms are same as in the NMSSM. For simplicity, we assume a common coupling $\lambda_V$ and corresponding soft breaking parameters for all of the  vector multiplets. The corresponding soft SUSY breaking terms are given by
\begin{eqnarray}
  \label{eq:soft-xy}
  -{\cal L}_{{\rm soft}}&=& M_X^2  |\widetilde{X}|^2 + M_Y^2  |\widetilde{Y}|^2
   +\left( \lambda_{V}A_{V}S\widetilde{X}\widetilde{Y}+B_{V}M_{V}\widetilde{X}\widetilde{Y}+\mathrm{H.c.}\right)
\end{eqnarray}
Then we have the Dirac fermion
$$V=\left(\begin{array}{c}
Y\\
\bar{X}\end{array}\right)$$
with $m_{V}=\lambda_{V}v_{s}$. The coupling of $h_{i}\bar{V}V$ interaction is proportional to $\lambda_{V}{\cal O}_{3i}$.  The mass matrix of the vector particle in the basis of $({\widetilde{X}}^\ast, \widetilde{Y})^{T}$ is given by (contribution from D-term have been ignored)
\begin{eqnarray}
  \label{eq:xy-sf}
\left(\begin{array}{cc}
\lambda_{V}^2 v_s^{2}+M_{X}^{2} &
\lambda_{V}\kappa v_s^{2}+\lambda_{V}A_{V}v_s+B_V M_V -\lambda\lambda_V v_u v_d\\
\lambda_{V}\kappa v_s^{2}+\lambda_{V}A_{V}v_s+B_V M_V -\lambda\lambda_V v_u v_d
& \lambda_{V}^2v_s^{2}+M_{Y}^{2}\end{array}\right)
\end{eqnarray}
After diagonalizing Eq.(\ref{eq:xy-sf})  to mass eigenstate $(\tilde{V}_1, \tilde{V}_2)^{T}$
by the rotation matrix $L$, we have the coupling
of $H_{i}\tilde{V}_{j}\tilde{V}_{k}^\ast$ in the mass eigenstates as
\begin{eqnarray}
  \label{eq:hvv}
\nonumber  &&2\lambda_{V}^2 v_s{\cal O}_{3i}(L^{2j}L^{2k\ast}+L^{1j}L^{1k\ast})
+\lambda_{V}(2\kappa v_s+A_{V}){\cal O}_{3i}(L^{2j}L^{1k\ast}+L^{1j}L^{2k\ast}) \\
&&-\lambda\lambda_V v_d {\cal O}_{1i}(L^{2j}L^{1k\ast}+L^{1j}L^{2k\ast})
-\lambda\lambda_V v_u {\cal O}_{2i}(L^{2j}L^{1k\ast}+L^{1j}L^{2k\ast})
\end{eqnarray}
In the end, the newly introduced free model parameters are $\lambda_{V},A_{V},B_V,M_{X}^{2},M_{Y}^{2},M_{V}^{2}$.

\section{numerical results and discussions}\label{sec3}
The new vector-like particles can contribute to the effective potential, which will significantly change the Higgs mass matrix and the loop-induced Higgs couplings, such as $hgg$ and $h\gamma\gamma$. At one-loop, their contributions are given by
\begin{eqnarray}
  V_1 &=& \sum_i \frac{n_iM^4_i}{64\pi^2}\left[\log\frac{M^2_i}{\Lambda^2}-\frac{3}{2}\right]
\end{eqnarray}
where $i$ denotes the mass eigenstate, $n_i$ depend on the color and spin of the mass eigenstate,
and $M^2_i$ are the field-dependent masses. We modify the
the package \textsf{NMSSMTOOLS} \cite{nmssmtools} to include the contributions of the vector-like particles.
Then we scan the parameter space of our model in the following range
\begin{eqnarray}
0.0<\lambda,\kappa<0.7,\ 1.0<\tan\beta<15.0,\ 100 {\rm GeV}<\mu<400 {\rm GeV},\ 0<A_{\lambda}<3 {\rm TeV}, \nonumber \\
|A_{\kappa}|< 1 {\rm TeV},\ 500 {\rm GeV}< m_{U_3,Q_3}<3 {\rm TeV},\ -3 {\rm TeV}<A_t<3 {\rm TeV},\ 400 {\rm GeV} <M_{1,2}<2 {\rm TeV}, \nonumber \\
0.0<\lambda_V<2,\ 0.1{\rm TeV} <M_X=M_Y=M_V<3 {\rm TeV},\ -4 {\rm TeV}<A_V,B_V<4 {\rm TeV}.
\end{eqnarray}
The soft mass parameters for the first two generation squarks and the sleptons are set to 2 TeV and other corresponding trilinear mass parameters are taken to zero. We also assume the gluino mass parameter $M_3=3$ TeV to satisfy the LHC bound. The generation of vector-like particles are chosen as 3. In the scan, we take the two singlet-like CP-even and CP-odd Higgs bosons masses in the range of 700~GeV-800~GeV for example and impose the following constraints:
\begin{itemize}
\item[(1)] We require the SM-like Higgs mass in the range of 123-127 GeV and use the 95\% exclusion limits from LEP, Tevatron and LHC in the Higgs searches with \textsf{HiggsBounds-4.2.0} \cite{higgsbounds}. We also perform the Higgs data fit by calculating $\chi^{2}$ of the Higgs couplings with the package \textsf{HiggsSignals-1.3.0} \cite{higgssignals} and require our samples to be consistent with the Higgs data at $2\sigma$ level.
\item[(2)] In order to obtain a stable color vacuum, the bilinear part need to be positively definite and the trilinear
terms cannot be too large:
\begin{eqnarray}
  |2M_VB_V| &<& M_X^2+M_Y^2 ,\\
  |A_V| &<& 2.5\sqrt{M_X^2+M_Y^2-|2M_VB_V|} .
\end{eqnarray}
\item[(3)] We require the thermal relic density of the lightest neutralino (as the dark matter candidate)
is lower than the upper bound of the Planck value \cite{planck}.
\item[(4)] We also consider the constraints from the null results of LHC searches for the dijet~\cite{jj}, diphoton~\cite{aa1,aa2} dibosons($ZZ,WW,hh$)~\cite{vv-atlas,vv-cms} and $t\bar{t}$ at the LHC Run-1.
\item[(5)] Since the LSP masses can be lighter than $m_{h_2/a_1}/2$ GeV, the Higgs bosons $h_2$ and $a_1$ can invisibly decay to the LSP, which will lead to the monojet signature $gg \to h_2/a_1(\to \chi^0_1 \chi^0_1) j$. We require our samples to satisfy the CMS monojet limit on the invisible decay~\cite{Khachatryan:2014rra} at 8 TeV LHC.
\end{itemize}
We calculate the production cross section of $gg \to h_2,a_1$ at the 13 TeV LHC by using the package
\textsf{HIGLU}\cite{higlu} with CTEQ6.6M PDFs \cite{cteq6}. We take the renormalization and factorization scales
as $\mu_R=\mu_F=m_S/2$. We also include a $K$-factor $(1+67\alpha_s/4\pi)$ \cite{qcdcorrection} in the calculation
of the decay width of $S \to gg$.

%%%Table 2 %%%%%%%%%%%%%%%%%%%%%%%
\begin{table}[h]
\caption{A benchmark point of degenerate heavy Higgs bosons in the NMSSM.}
\begin{tabular}{|  c | c | c|c|c|c|c|c|}
\hline
 $\lambda$ & $\kappa $ & $\tan \beta$ &  $\mu$ (GeV) &$A_\lambda$ (GeV) &$A_\kappa$ (GeV) &$m_{Q3}$ (GeV)\\
\hline
 0.594&0.428&14.65&113.5&1908.5&-631.5&1855\\
\hline
$m_{U3}$ (GeV) & $A_t$ (GeV) &$M_1$ (GeV) & $M_2$(GeV) & $N_f $ & $\lambda_V$ & \\
\hline
 1997&-2693.6&1984.81&512.19&3&5.06&\\
\hline
$M_X$ (GeV) &$A_V$ (GeV) &$B_V$ (GeV) & $m_{h1}$ (GeV) &$m_{h2}$(GeV) & $m_{h3}$(GeV)&\\
\hline
 427.3&307.1&-96.12&124.36&745.7&1448.3&\\
\hline
$m_{a1}$(GeV) & $m_{a2}$(GeV) & $m_{h\pm}$(GeV)& $m_{\chi^{\pm}_1}$(GeV)&$m_{\chi^{0}_1}$(GeV)&$m_{\chi^{0}_2}$(GeV)&$m_{\chi^{0}_3}$(GeV)\\
\hline
 750.6&1775.0&1770.8&112.79&70.3&141.7&226.8\\
\hline
$m_{V_F}$(GeV) &$m_{V_{s_1}}$(GeV)&$m_{V_{s_2}}$(GeV)&$\mathcal{O}^2_{13}$&$\Omega h^2$&$\sigma_{gg\rightarrow h_2}^{13TeV}$(fb)&$\Gamma(h_2)$(GeV)\\
\hline
 968.8&893.1&1202.09&$2.79\times 10^{-5}$&0.0906&6885.8&12.8\\
\hline
$BR_{h_2\rightarrow \tau\tau}$ &$BR_{h_2\rightarrow bb}$&$BR_{h_2\rightarrow tt}$&$BR_{h_2\rightarrow WW}$&$BR_{h_2\rightarrow ZZ}$&$BR_{h_2\rightarrow \gamma\gamma}$&$BR_{h_2\rightarrow Z\gamma}$\\
\hline
 0.025\%&0.174\%&0.0006\%&0.026\%&0.014\%&0.022\%&0.013\%\\
\hline
$BR_{h_2\rightarrow gg}$&$BR_{h_2\rightarrow h_1 h_1}$&$BR_{h_2\rightarrow \chi^{0}_1 \chi^{0}_1}$&$BR_{h_2\rightarrow \chi^{0}_1\chi^{0}_2}$&$BR_{h_2\rightarrow \chi^{0}_1 \chi^{0}_3}$&$BR_{h_2\rightarrow \chi^{0}_2 \chi^{0}_2}$&$BR_{h_2\rightarrow \chi^{0}_2 \chi^{0}_3}$\\
\hline
 6.8\%&0.059\%&22.7\%&4.6\%&0.98\%&11.8\%&6.1\%\\
\hline
$BR_{h_2\rightarrow \chi^{0}_3 \chi^{0}_3}$&$BR_{h_2\rightarrow \chi^{\pm}_1 \chi^{\pm}_1}$&$\sigma_{gg\rightarrow a_1}^{13TeV}$(fb)&$\Gamma(a_1)$(GeV)&$BR_{a_1\rightarrow \tau\tau}$ &$BR_{a_1\rightarrow bb}$&$BR_{a_1\rightarrow tt}$\\
\hline
 14.3\%&31.6\%&7689.14&14.6&0.0098\%&0.068\%&0.0049\%\\
\hline
$BR_{a_1\rightarrow WW}$&$BR_{a_1\rightarrow ZZ}$&$BR_{a_1\rightarrow \gamma\gamma}$&$BR_{a_1\rightarrow Z\gamma}$&$BR_{a_1\rightarrow gg}$&$BR_{a_1\rightarrow h_1 Z}$&$BR_{a_1\rightarrow \chi^{0}_1 \chi^{0}_1}$\\
\hline
 0&0.0018\%&0.022\%&0.013\%&6.7\%&$9.5\times 10^{-8}$&16.9\%\\
\hline
$BR_{a_1\rightarrow \chi^{0}_1 \chi^{0}_2}$&$BR_{a_1\rightarrow \chi^{0}_1 \chi^{0}_3}$&$BR_{a_1\rightarrow \chi^{0}_2 \chi^{0}_2}$&$BR_{a_1\rightarrow \chi^{0}_2 \chi^{0}_3}$&$BR_{a_1\rightarrow \chi^{0}_3 \chi^{0}_3}$&$BR_{a_1\rightarrow \chi^{\pm}_1 \chi^{\pm}_1}$&$ \sigma^{13TeV}_{\gamma\gamma}$(fb)\\
\hline
 4.3\%&0.27\%&10.8\%&6.38\%&22.3\%&31.4\%&3.22\\
\hline
\end{tabular}
\label{bp1}
\end{table}

In Table \ref{bp1}, we present a benchmark point that can successfully explain the diphoton excess under the
constraints (1)-(5). From this table we can see:
\begin{itemize}
  \item The lightest CP-even Higgs boson $h_1$ is SM-like.
  \item The total decay widths of the CP-even and CP-odd singlet-like Higgs bosons are 12.8 GeV and 14.6 GeV, respectively.
  \item The dominant decay mode of $h_2$ is $h_2 \to gg$, while $a_1$ mainly decays to $\chi^+_1 \chi^-_1$.
  \item The production cross sections of $gg \to h_2$ and  $gg \to a_1$  can reach 6.9 pb and 7.7 pb, respectively.
             The branching ratios of $h_1 (a_1) \to \gamma\gamma$ are about $0.022\%$. Then, the total production rate
             of the diphoton can be 3.22 fb.
\end{itemize}

%%%% Fig.1 %%%%%%%%%%%%%%%%%%%%%%%%%%%%%%
\begin{figure}[ht]
\centering
\includegraphics[width=12cm]{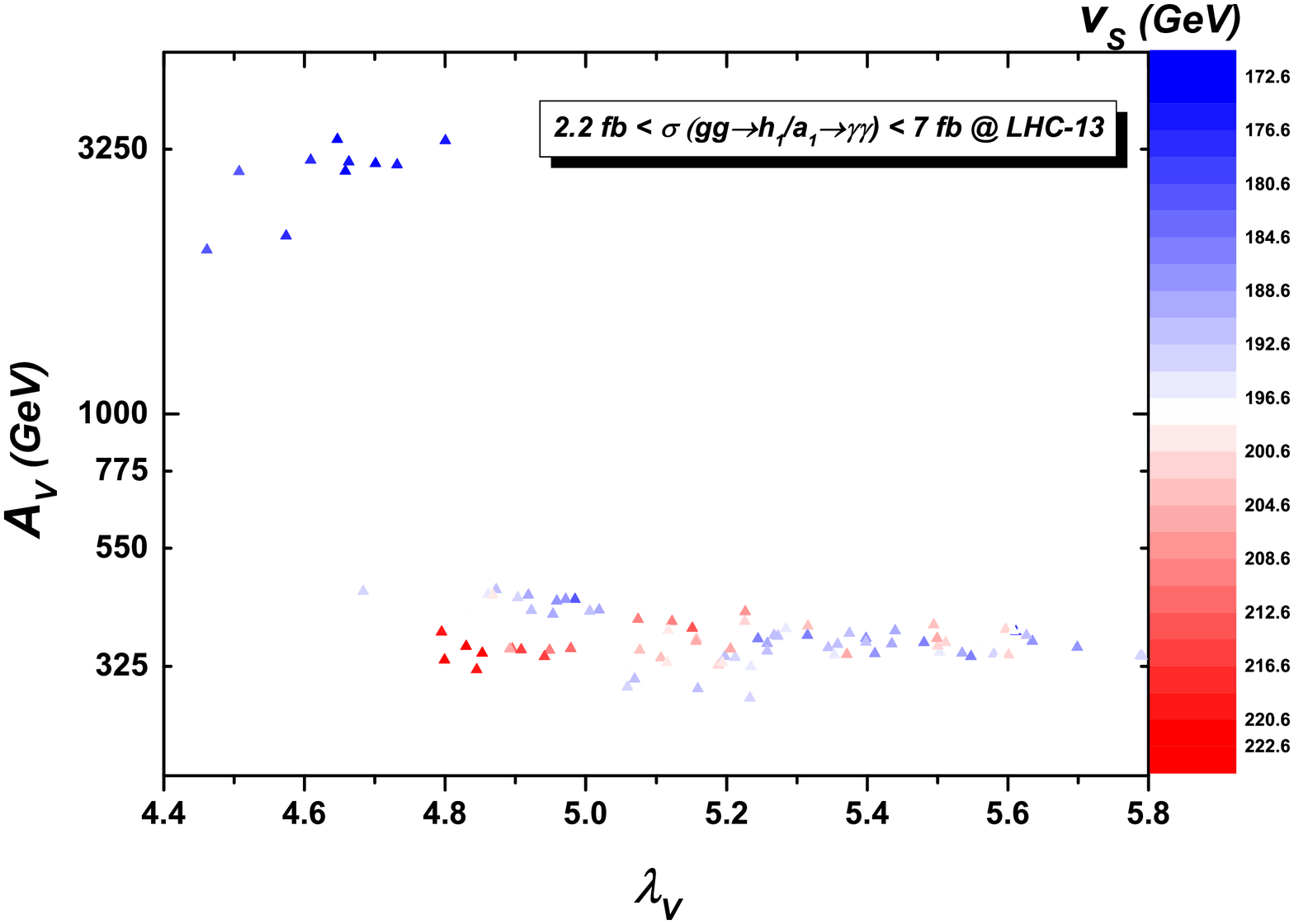}
\vspace{-0.5cm}
\caption{Scatter plots of the survived samples on the plane of $\lambda_V$ versus $A_V$.
The color map corresponds to the values of $v_S$. All the samples satisfy the constraints (1)-(5).}
\label{re1}
\end{figure}
%%%%%%%%%%%%%%%%%%%%%%%%%

In Fig.\ref{re1}, we plot the surviving samples on the plane of $\lambda_V$ versus $A_V$. All the samples satisfy the
constraints (1)-(5). From Fig.\ref{re1} we can see that when the Yukawa coupling
$\lambda_V$ becomes smaller, the trilinear parameter $A_V$ should be larger and the higgsino mass $\mu$ tends to be lighter.
Otherwise, the large $\lambda_V$ and $A_V$ will overly enhance the diphoton production rate and are disfavored the LHC data. The favorable ranges of $\lambda_V$ and $v_S$ are $4.4-5.8$ and $170 {\rm GeV}-225 {\rm GeV}$, respectively.

%%%% Fig.2 %%%%%%%%%%%%%%%%%%%%%%%%%%%%%%
\begin{figure}[ht]
\centering
\includegraphics[width=12cm]{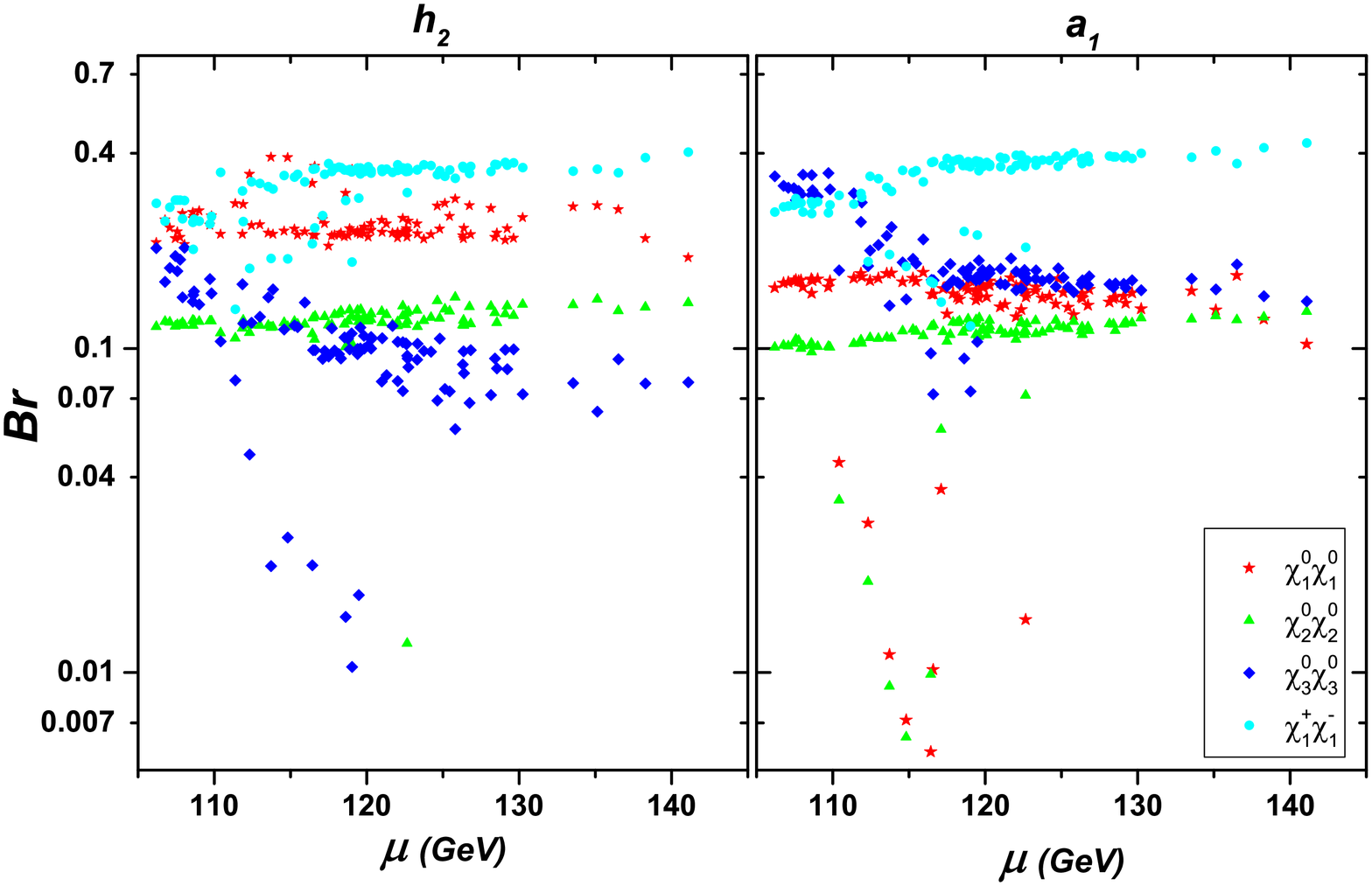}
\vspace{-0.5cm}
\caption{Same as Fig.\ref{re1}, but showing the decay branching ratios of $h_2$ (left panel) and $a_1$ (right panel)
versus the higgsino mass parameter $\mu$.}
\label{re2}
\end{figure}
%%%%%%%%%%%%%%%%%%%%%%%%%

Due to the light LSP and higgsinos, the two degenerate heavy Higgs bosons $h_2$ and $a_1$ can decay to $\chi^0_{1,2,3}$
and $\chi^+_1 \chi^-_1$. In Fig.~\ref{re2} we present the decay branching ratios of $h_2$ and $a_1$. It can be seen that
both $h_2$ and $a_1$ dominantly decay to $\chi^+_1 \chi^-_1$ with a branching ratio $Br \simeq 40\%$. The invisible decay
branching ratio of $h_2(a_1) \to \chi^0_1 \chi^0_1$ can maximally reach about $30\%$($20\%$). Thus, the future search
for the monojet events through the process $gg \to h_2 / a_1 (\to \chi^0_1 \chi^0_1) j$ can further test our scenario.

\section{Conclusions}\label{sec4}
We studied the degenerate heavy Higgs bosons in the NMSSM by introducing vector-like particles. In our model, the decays of $h_2(a_1) \to \gamma\gamma$ are dominated by the vector-like squarks with small singlet vev $v_S$, which can also enhance the production cross section of $gg \to h_2/a_1$. Under the current LHC constraints and the dark matter detection limits, we scanned the parameter space and found that such a scenario has been stringently constrained by the null results of the LHC searches for heavy mass resonances. We also noticed that the two heavy Higgs bosons $h_2$ and $a_1$ have sizable invisible decay branching ratio $h_2(a_1) \to \chi^0_1\chi^0_1$. Therefore, the future search for the monojet events through the process $gg \to h_2 / a_1 (\to \chi^0_1 \chi^0_1) j$ can further test our scenario at the LHC.

\section*{Acknowledgments}
This work is partly supported by the Australian Research Council, by the National Natural Science Foundation of China (NNSFC)
under grants Nos. 11705093 and 11675242, by the Open Project Program of State Key Laboratory of Theoretical Physics,
Institute of Theoretical Physics, Chinese Academy of Sciences (No.Y5KF121CJ1); by the Innovation Talent project of Henan Province under grant number 15HASTIT017; by the Young-Talent Foundation of Zhengzhou University; by Ri-Xin Foundation of BJUT,
and by the CAS Center for Excellence in Particle Physics (CCEPP).

\end{document}